\documentclass[12pt]{iopart} % style of IOP Journals%

\usepackage{iopams}
\usepackage{setstack}
\usepackage{graphicx}
\usepackage{amsfonts}

\newcommand{\BEA}{\begin{eqnarray}}
\newcommand{\EEA}{\end{eqnarray}}

\usepackage{mathrsfs}
\usepackage{color}

\newcommand{\abs}[1]{\left| #1\right|} % absolute value
\newcommand{\ket}[1]{\left|#1\right>} % ket
\begin{document}

\title{All photons are equal but some photons are more equal than others}
\author{Falk T\"{o}ppel$^{1,2}$, Andrea Aiello$^{1,2}$ and Gerd Leuchs$^{1,2}$}
\author{}
\ead{falk.toeppel@mpl.mpg.de} %corresponding author
\address{$^1$ Max Planck Institute for the Science of Light, G\"{u}nther-Scharowsky-Stra{\ss}e 1/Bldg. 24, 91058 Erlangen, Germany}
\address{$^2$ Institute for Optics, Information and Photonics, Universit\"{a}t Erlangen-N\"{u}rnberg, Staudtstra{\ss}e 7/B2, 91058 Erlangen, Germany}
\date{\today}

\begin{abstract}
%An operational approach to the problem of photons distinguishability  is presented.  
Two photons  are said to be identical when they are prepared in the same quantum state. Given the latter, there is a unique way to achieve this.
Conversely, there are many different manners to prepare two non-identical photons: they may have different frequency, polarization, amplitude, etc.
Therefore, photon distinguishability depends upon the specific degree of freedom being varied.
By means of a careful analysis of the coincidence probability distribution in a Hong-Ou-Mandel experiment, we can show that photon distinguishability 
can be actually quantified by the rate of distinguishability of photons, an experimentally measurable parameter that crucially depends on both the photon quantum state and the degree of freedom under control.
\end{abstract}

\pacs{03.65.Ta, 03.65.Ca, 03.70.+k, 14.70.Bh}
\submitto{\NJP}

\maketitle

\section{Introduction}
Scalable implementations of many promising linear optics quantum computation (LOQC) schemes require repeated occurrence of two-photon interference effects \cite{KLM,Enk11,Filip11}. In these protocols individual photons must be carefully prepared in two distinct optical modes as, e.g., TE or TM polarization modes \cite{NC} and HG$\/_{01}$ or HG$\/_{10}$ spatial modes \cite{Souza}, in order to implement bona fide qubits. Arbitrary mode control of a single photon has been recently demonstrated for photon's amplitude \cite{Keller,McKeever,Kuhn,Bochmann,Kolchin}, polarization \cite{Wilk}, frequency \cite{Kuhn04} and phase \cite{Specht09}. This variety of results leads to the question: are all these distinct degrees of freedom (polarization, frequency, etc.) equivalent in determining two-photon interference? As perfect interference requires identically prepared photons, the question above can be rephrased as: how the control of a specific degree of freedom (DOF) affects photon distinguishability? 

In this paper we answer to this question by introducing in an operational manner the concept of \emph{rate of distinguishability} of photons.  This parameter permits to quantify the effects upon photon distinguishability of the variation of a single, arbitrary DOF and it can be actually measured in a two-photon interference experiment. Our results suggest the need for replacement of the  strong concept of ``photon indistinguishability'',  with the weaker concept of ``photon indistinguishability with respect to a given degree of freedom''.

\section{Two-photon interference}
When two equally prepared photons interfere at the two input ports of a $50/50$ beam splitter (BS), the joint probability of detection (coincidence probability) at the two outputs is exactly zero. This phenomenon is known as photon \emph{coalescence}. Vice versa, when the two photons are prepared in a different manner, the coincidence probability raises up to $50 \%$.  This effect was first demonstrated by Hong \emph{et al.}  \cite{Hong87} and  rapidly became central in a broad range of experiments in quantum physics \cite{Kaltenbaek09,Solomon10}. 
A typical experimental layout is sketched in figure 1.
%
%\vspace{7.5truecm}
%
%%%%%%%%%%%%%%%%%%%%%%%%%%%%%%%%%%%%%%%%%%%%%%%%%%%%%%%%%%%%%%%%%%%%%%%%%%%%%%%%%%
%
\begin{figure}[h!]
\begin{center}
\includegraphics[angle=0,width=7.0truecm]{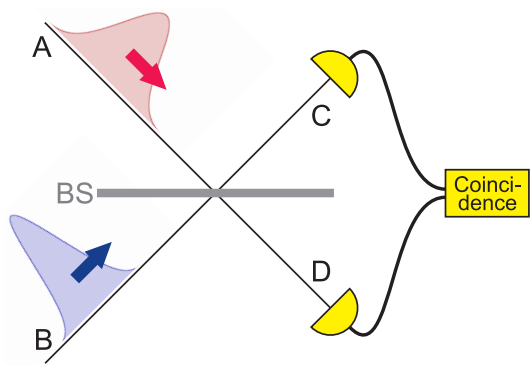}
\caption{\label{fig:1} (color online) Two-photon interference at a $50/50$  beam splitter. Two independently prepared photons enters the two input ports $A$ and $B$ of the BS. They are eventually detected by two distinct detectors placed behind the output ports  $C$ and $D$ of the BS. The plane of the figure is the \emph{plane of incidence}}
\end{center}
\end{figure}
%
%%%%%%%%%%%%%%%%%%%%%%%%%%%%%%%%%%%%%%%%%%%%%%%%%%%%%%%%%%%%%%%%%%%%%%%%%%%%%%%%%%%
%
%
In the present work, the two \emph{independent} photons impinging upon the $50/50$ BS are prepared in the product state $| \Psi^{AB} \rangle = | \Psi^A \rangle | \Psi^B\rangle$, where
$| \Psi^A \rangle  = \hat{a}^\dagger[\psi^A]| 0 \rangle$ ($| \Psi^B \rangle  = \hat{b}^\dagger[\psi^B]| 0 \rangle$)
denotes the single-photon state in arm $A$ ($B$), with $\hat{a}^\dagger[\psi^A]$ ($\hat{b}^\dagger[\psi^B]$) being the operator that creates one photon in the    wavepacket  mode (or, simply, wave function) $\psi^A$ ($\psi^B$) \cite{Deutsch91,Loudon98}.\footnote{Throughout this paper we will use capital and lower case Greek letters to denote a \emph{state vector}, say $\ket{\Psi}$, and the corresponding \emph{wave function}, say $\psi$, respectively \cite{MerzbacherBook}. } The \emph{normalized} wave functions  $\psi^A$ and $\psi^B$ completely fix the spectral, polarization and spatiotemporal characteristics of the photon entering port $A$ and $B$, respectively. 

%We focus our attention on two-photon pure states instead of more general mixed states. The main reason for this is that we are interested in the study of the fundamental indistinguishability properties of photons. Mixedness of a quantum state is a consequence of the subjective classical ignorance of the experimenter in the state preparation, rather than an intrinsic quantum property of the photon. Hence, such a classical ignorance is not unavoidable and could, in principle, be removed. However, the extension of our formalism to mixed states is straightforward.

Annihilation and creation operators associated to orthogonal wave functions, do commute: $\big[ \hat{a}[\psi],\hat{b}^\dagger[\phi]\big] = \left( \psi,\phi \right)\delta_{a b} $, where $\left( \psi,\phi \right)$ denotes the scalar product in the complex linear space of the wave functions $\mathscr{L} \ni \psi, \phi$ \cite{KolmogBook}.
The probability of detecting the two photons at the two output ports $C$ and $D$ is given by  \cite{LoudonBook}:
\BEA\label{eq10}
P_{1,1}[\psi^A,\psi^B] = \left( 1 - \abs{ \langle \Psi^A | \Psi^B \rangle }^2\right)/2,
\EEA
where $ \langle \Psi^A | \Psi^B \rangle = \left( \psi^A,\psi^B\right)$.

Now, assume that the  two input photons are prepared ``almost''  in the same manner, in such a way that they can be represented by the wave functions
%
%single-photon wave packet modes $\psi^A$ and $\psi^B$ are prepared ``almost''  in the same manner, namely 
$\psi^A = \psi$ and $\psi^B  = \left( \psi+  \delta \psi \right)/ \| \psi+  \delta \psi \| \equiv \widetilde{\psi} + \widetilde{\delta \psi} $.
The  functional variation of the  coincidence probability generated by $\delta \psi$ will be, by definition: $\Delta P_{1,1}[\psi]\equiv P_{1,1}[\psi, \widetilde{\psi} + \widetilde{\delta \psi}] - P_{1,1}[\psi, \psi ]=  P_{1,1}[\psi, \widetilde{\psi} + \widetilde{\delta \psi}]$, where $P_{1,1}[\psi, \psi] =0$ for identically prepared photons, as it trivially follows from \eref{eq10} and normalization $\langle \Psi | \Psi \rangle =(\psi,\psi)=1$.
 A straightforward calculation from  \eref{eq10} yields
\BEA\label{eq20}
\Delta P_{1,1}[\psi] =\frac{1}{2} \frac{\Delta^2\left( 1 - |\alpha|^2\right)}{ 1 + \Delta (\alpha + \alpha^*) + \Delta^2},
\EEA
where $\Delta^2 \equiv (\delta \psi, \delta \psi)$, $\alpha  \equiv (\psi, \delta \psi)/ \Delta$ with $|\alpha|<1$ and $(\psi,\psi) = 1$. This result is \emph{exact} and
rests on  the basic properties of the scalar product in a complex linear space $\mathscr{L}$ solely.

Equation \eref{eq20} may be further developed by assuming that the functional deviation $\delta \psi$ is generated by the variation of
 a single DOF, represented by the \emph{real} parameter $f$, in such a way that $\psi^A = \psi(f)$ and $\psi^B = \psi(f+  \delta\!f)$, with $|\delta\!f| \ll |f|$ and $\bigl(\psi(f),\psi(f)\bigr)=1$ for all $f$. For example, if $f=\nu$, the photon at input port $A$ has central frequency $\nu^A = \nu$ and the one entering port $B$ has central frequency $\nu^B = \nu + \delta \nu$.
Defining  $\delta \psi \equiv \psi(f + \delta\!f) - \psi (f)$ permits to express $\psi^B$ as above: $\psi^B  = \psi+  \delta \psi \equiv \widetilde{\psi} + \widetilde{\delta \psi} $.
Formally expanding $\psi(f+  \delta\!f)$ in powers of $\delta\!f$ as  \footnote{The expansion is formal in the sense that we \emph{assume} the existence and the continuity of first and second-oder derivatives $\psi'$ and $\psi''$ respectively. If this condition fails then our theory may become not applicable.}
\BEA\label{eq30}
\psi(f+  \delta\!f) =\exp\left(\delta f\frac{\partial}{\partial f}\right) \psi(f)\simeq \psi(f)+  \psi'(f) \, {\delta  f} + \psi''(f) \,\frac{{\delta  f}\/^{\,2}}{2}  + \dots,
\EEA
with $\psi'(f) = \partial \psi(f)/ \partial f$, $ \psi''(f) =\partial^2 \psi(f)/ \partial f^2$, et cetera,  we can straightforwardly obtain %
\BEA\label{eq40}
\Delta^2 = {\delta\!f}^2 \Bigl[\left(\psi', \psi'\right)+ \mathrm{Re} \left[\left(\psi'',\psi' \right)\right] \delta\!f + \Or({\delta\!f}^2) \Bigr],
\EEA
and $\alpha = \left[\left(\psi, \psi'\right) + \left( \psi, \psi'' \right) \delta\!f /2  + \Or({\delta\!f}^2)\right]/\Delta$. Since from \eref{eq20} and \eref{eq40} it follows that $\Delta P_{1,1}[\psi] \propto \Delta^2 = \Or({\delta\!f}^2)$, we define the \emph{rate of distinguishability} $R_{f}[\psi]$ of the photons with respect to the degree of freedom $f$ via the relation
\BEA\label{eq60}
R_{f}[\psi] &\equiv  {\left.\frac{\partial^2}{\partial\,\delta\!f^2}  P_{1,1}[\psi(f),\psi(f + \delta\!f)] \right|_{\delta\!f=0}}  \nonumber \\
&=\left(\psi', \psi' \right) - |\left(\psi, \psi'\right)|^2  ,
\EEA
where $0 \leq R_{f}[\psi] \leq  \| \psi' \|^2$, and $\left(\psi, \psi'\right)^2 \leq 0$ because $0 = \partial\left(\psi, \psi\right)/\partial f = \left(\psi, \psi'\right) + \left(\psi', \psi\right)$. 

When defining the generator of a translation in the parameter $f$ as $\hat{K}=-\rmi \partial/\partial f$, one may understand the Taylor expansion \eref{eq30} in terms of a propagator: $\psi(f+\delta\!f)=\exp(\rmi \delta\!f \hat{K})\psi(f)$. In general, $(\hat{K}\psi,\phi)\neq(\psi,\hat{K}\phi)$ for arbitrary wave functions $\psi$ and $\phi$  because $f$ is just a parameter upon which the photon wave function depends and \emph{not} a dynamical variable. For this reason, the operator $\hat{K}$ is, in general, not self-adjoint. However, for some DOFs and certain states, e.g., spatial displacement of Gaussian states in wave vector representation, the relation $\left(\psi', \psi'\right)=-\left(\psi, \psi''\right)$ holds. In these cases the rate of indistinguishability $R_{f}[\psi]$ simplifies to the variance of $\hat{K}$, and we obtain a link to the geometry of quantum states \cite{Anandan}. \footnote{We are thankful to an anonymous Referee for pointing out this connection.}

The  dimensionless parameter $R_{f}[\psi] \delta\!f^2$  has a straightforward physical meaning: it tells us how rapidly two identically prepared photons become distinguishable when we slightly vary, from $\psi(f)$ to $\psi(f + \delta\!f)$, the wave function of one photon with respect to the other.  %The parameter $R_{f}[\psi]$ has some interesting properties. 
Thus, given a pair of photons prepared in the same state $|\Psi \rangle$, one can assert that they are \emph{maximally indistinguishable} with respect to $f$ if  $R_{f}[\psi] \leq R_{\bar{f}}[\psi]$ for any possible degree of freedom $\bar{f}$. In a complementary manner, provided two distinct pairs of photons, the first two photons being prepared in the state $|\Psi\rangle$ and the second ones in the state $|\Phi\rangle$, one can say that the photons in the first pair are maximally indistinguishable with respect to $\psi$ for a fixed $f$, if $R_{f}[\psi] \leq R_{f}[\phi]$ for all possible wave functions $\phi$. In this case it is not difficult to prove that the rate of distinguishability is minimal for a Gaussian shaped wave function \cite{Rohde}.
In summary:
 ``all photons are equal but some photons are more equal than others'' \footnote{Freely adapted from: Orwell G 2008 {\it Animal Farm} (London: Penguin Books Ltd) chapter X.}, and  $R_{f}[\psi]$ quantifies the degree of equality. 

This result concludes the first part of this work. Next, we will apply equations \eref{eq30} and \eref{eq40} to the realistic case of  optical Gaussian wave packets with well-defined spectral, spatiotemporal and polarization DOFs. %We will find that coupling between different DOFs may heavily alter the rate of distinguishability of photons.

\section{Gaussian wave packets.}
Consider a single-photon wave packet of the form:
\BEA\label{eq:quant_state_gen}
\ket{\Psi}=\sum_{s=1}^2{\int} \rmd^3k\,\psi_s(\bi{k})\hat{a}^\dagger_s(\bi{k})\ket{0},
\EEA
where $\ket{0}$ denotes the ground state of the continuous Fock space and $\hat{a}_s(\bi{k})$ is the operator that annihilates one photon from the plane wave mode $\bi{e}_s(\bi{k})\exp \left( \rmi\bi{k}\cdot\bi{r} \right)$, with $\bigl[ \hat{a}_s(\bi{k}), \hat{a}^\dagger_{s'}(\bi{k}')\bigr] = \delta^{(3)}(\bi{k}-\bi{k}') \delta_{s s'}$ \cite{MandelBook}. Here $\left\{\bi{e}_1(\bi{k}),\bi{e}_2(\bi{k}),\bi{k}/\abs{\bi{k}} \right\}$ denotes a right-handed orthonormal basis set attached to the wave vector $\bi{k}$.
The normalization of the state is ensured by requiring $\langle \Psi | \Psi \rangle = \sum_{s=1}^2{\int} \rmd^3k\,|\psi_s(\bi{k})|^2=1$.
%
%\BEA\label{eq:normalisation_state_fkt}
%
%\langle \psi | \Psi \rangle = \sum_{s=1}^2{\int} \rmd^3k\,|\psi_s(\bi{k})|^2=1.
%
%\EEA
%
The spectral amplitudes $\psi_s(\bi{k})$ ($s=1,2$) determine the shape and the polarization of the beam. They may be obtained by imposing the quantum-classical correspondence 
\BEA\label{eq80}
\langle 0| \hat{\bi{E}}^{(+)}(\bi{r},t)| \Psi \rangle = \bi{E}^{(+)}_{\rm{cl}}(\bi{r},t),
\EEA
where $\bi{E}^{(+)}_{\rm{cl}}(\bi{r},t)$ is the positive-frequency part of the classical field wave packet whose energy  equals the mean energy of the photon in the state $| \Psi \rangle$, and
\begin{equation}\label{eq90}
\eqalign{
\hat{\bi{E}}^{(+)}(\bi{r},t) =& \frac{\rmi}{(2 \pi )^{3/2}}\sum_{s=1}^2
 \int \rmd^3 k \, \sqrt{\frac{\hbar \omega}{2 \epsilon_0}}\hat{a}_s(\bi{k}) \bi{e}_s(\bi{k}) \exp \left[\rmi \left( \bi{k} \cdot \bi{r} - \omega t\right) \right],}
\end{equation}
with $\omega = c \abs{\bi{k}} \equiv c k $, $c$ being the speed of light in vacuum and $\epsilon_0$ the vacuum permittivity. The expression for $\bi{E}^{(+)}_{\rm{cl}}(\bi{r},t)$ is given by the right side of  \eref{eq90} with the quantum operator $\hat{a}_s(\bi{k})$ replaced by the classical amplitude $\tilde{a}_s(\bi{k})$. Then, by substituting from equations \eref{eq:quant_state_gen} and \eref{eq90} into  \eref{eq80}, one attains $\psi_s(\bi{k})=\tilde{a}_s(\bi{k})$. The total energy contained in such wave packet is given by $\mathcal{E} = \int \rmd^3 k  \, \hbar \omega \left( \abs{\tilde{a}_1(\bi{k})}^2 + \abs{\tilde{a}_2(\bi{k})}^2 \right)$.

Without loss of generality, we assume that $\tilde{a}_s(\bi{k}) = \varepsilon_s(\bi{k})E(\bi{k})$, where $E(\bi{k})$ and $\varepsilon_s(\bi{k})$ are the scalar and the vector spectral amplitudes of the field, respectively. $E(\bi{k})$ determines the spatial characteristics of the field, and $\varepsilon_s(\bi{k})$ the polarization ones. Here we consider a collimated, quasi-monochromatic wave packet, with central wave vector  $\bi{k}_0$ and central frequency $\omega_0 = c \abs{\bi{k}_0} \equiv c k_0$. We choose
a normalized Gaussian spectral amplitude $E(\bi{k}) = \gamma(\bi{k} - \bi{k}_0)$, where
\BEA
\label{eq:Gaussian_f}
 \gamma(\bi{q}) = \frac{ (\det V)^{1/4}}{\pi^{3/4} } \exp \left[ -\rmi\;\bi{q}\cdot\bi{r}_0 - \frac{1}{2}\; \bi{q}\cdot  V \bi{q} \right] ,
\EEA
with $V^{-1} = \mathrm{diag} \left( \sigma_1^2, \sigma_2^2, \sigma_3^2 \right)$.  
This choice for $V$ yields a factorizable spectral amplitude $\gamma(\bi{q}) = g(q_1) g(q_2) g(q_3)$ with $g(q_n) = \exp[-\rmi q_n r_{0n}-q_n^2/(2 \sigma_n^2)]/(\pi^{1/4}\sqrt{\sigma_n})$. Clearly, it is possible to consider a more general positive definite symmetric matrix $V$ that couples different wave vector coordinates. We will see later that such a coupling may have dramatic consequences upon the rate of distinguishability of photons. 
In  \eref{eq:Gaussian_f} the real vector $\bi{r}_0 = \{r_{01},r_{02},r_{03} \}$ gives the position, at time $t=0$, of the center of the wave packet.
We fix $\varepsilon_s(\bi{k})$ assuming that the wave packet has passed across a polarizer that selects a \emph{uniform} field polarization parallel to  $\bi{p} \in \mathbb{C}^3$ and perpendicular to $\bi{k}_0$, with $\abs{\bi{p}}^2 = 1$ and  $\bi{k}_0\cdot\bi{p}=0$. In this case, it becomes natural to define $\varepsilon_s(\bi{k})$ as the normalized projection of $\bi{p}$ upon $\bi{e}_s(\bi{k})$, namely \cite{FainmanANDShamir,Aiello:09}: $\varepsilon_s(\bi{k})={\bi{e}_s(\bi{k})\cdot\bi{p}}/{\sqrt{1-|(\bi{p},\bi{k})|^2/k^2}}$,
with $\left|\varepsilon_1(\bi{k})\right|^2 + \left|\varepsilon_2(\bi{k})\right|^2=1$ by definition.

The Gaussian distribution $\gamma(\bi{k}-\bi{k}_0)$ implies that the wave packet is concentrated in a region of the $\bi{k}$-space of ``volume'' $\sigma_1 \sigma_2 \sigma_3$ centered at $\bi{k}_0$. Then, the assumptions of collimation and quasi-monochromaticity entail the constraints $\sigma_i \ll k_0, \; (i = 1,2,3)$. In this case, the total energy of the wave packet can be written as $\mathcal{E} = \int \rmd^3 k  \, \hbar \omega \abs{ \gamma(\bi{k}-\bi{k}_0)}^2 \simeq \hbar \omega_0$, where  $\int \rmd^3 k  \, \abs{ \gamma(\bi{k}-\bi{k}_0)}^2 =1$ by definition.

For quasi-monochromatic and collimated beams the Gaussian spectral amplitude $\psi_s(\bi{k}) = \varepsilon_s(\bi{k}) \gamma(\bi{k}-\bi{k}_0)$ contains $(3+3)+3 + 3 = 12$ independent real parameters corresponding to the (spectral) $\oplus$ spatial $\oplus$ polarization DOFs: $ \left( \bi{k}_0 \, \oplus\{ \sigma_1, \sigma_2, \sigma_3 \} \right) \, \oplus \bi{r}_0 \, \oplus \{ \bi{p} \in \mathbb{C}^3: \abs{\bi{p}}^2 = 1 \, \wedge \, \bi{k}_0\cdot\bi{p}=0 \}$.
Note that the central frequency $\omega_0$ is \emph{not} an additional independent parameter, since $\omega_0 = c \abs{\bi{k}_0}$.  %Thus, there are $12$ (actually $15$ if we consider a non-diagonal symmetric $V$), independent parameters to characterize  the quantum state of the photon.
Each of these $12$ (actually $15$ if we consider a non-diagonal symmetric $V$) parameters can be taken as the variable $f$  to evaluate the rate of distinguishability $R_f[\psi]$. This calculation will be the goal of the remainder.

\section{Rate of distinguishability}
Using rather standard methods of calculation \cite{MonkenPRL,Wang}, it is not difficult to show
that the coincidence probability \eref{eq10} can be expressed in terms of the spectral amplitudes $\psi_s^A(\bi{k})$ and $\psi_s^{B}(\bi{k})$ of the input photons as
\BEA\label{eq100}
\!\!\!P_{1,1}[\psi^A, \psi^B]=\frac{1}{2}\!\left[1-\biggl|\sum_{s=1}^2{\int} \rmd^3k\,{\psi_s^A(\bi{k})\psi_s^{B}}^*(\underline{\bi{k}})\biggr|^2\right],
\EEA
where $\underline{\bi{k}}$ has components $\{-k_1,k_2,k_3\}$. This change of sign in the $1$-coordinate is due to the parity inversion occurring by reflection at the BS. Hereafter, we assume two Gaussian wave packets $\psi_s^A(\bi{k}) = \psi_s(\bi{k}, f)$ and $\psi_s^B(\bi{k}) = \psi_s(\bi{\underline{k}\,}, f + \delta\!f)$. Moreover, for concreteness, we choose the $3$-axis of the Cartesian reference frame directed along $\bi{k}_0$, namely $\bi{k}_0 = \{0,0,k_0  \}$.

The explicit values of $R_f$, calculated from  \eref{eq60}, are given in table 1  below, for spectral and spatial DOFs:

\begin{table}[h!]
\caption{Rate of distinguishability $\displaystyle{R_f}$ for several spectral and spatial degrees of freedom $f$ of the photons, with $n=1,2,3$.}
\begin{indented}
\item[]\begin{tabular}{@{}lccc}
\br
~~\qquad\quad$f$\quad\qquad & \qquad\quad$k_{0n}$\quad\qquad & \qquad\quad$\sigma_n$\quad\qquad & \qquad\quad$r_{0n}$\quad\qquad \\
\mr
~\qquad\quad$\displaystyle{R_f}$\qquad & $\displaystyle{\frac{1}{2 \sigma_n^2}}$ &
$\displaystyle{\frac{1}{2 \sigma_n^2}}$ & $\displaystyle{\frac{\sigma_n^2}{2}}$\\
\br
\end{tabular}
\end{indented}
\end{table}
A remarkable consequence from table 1, is that for the complementary position$/$wave-vector variables, the following Fourier-transform equality holds: \footnote{We conjecture without demonstration that for non-Gaussian wave packets the minimum-uncertainty equality \eref{eq:Prod} will be replaced by $R_{k_{0n}}R_{r_{0n}} \geq {1}/{4}$.}
\BEA
\label{eq:Prod}
R_{k_{0n}}R_{r_{0n}} = {1}/{4} , \qquad \forall \, n=1,2,3 .
\EEA
table 1 furnishes some valuable information. Consider, for example, the last column: it  shows that  $R_{r_{0n}}^{1/2} {\delta r_{0n}}$ is equal to the ratio  between the variation  $\delta r_{0n}$ and the \emph{standard deviation} (square root of the variance) $\sqrt{2}/\sigma_n$ of the absolute value squared of the photon wave function in configuration space. This is in agreement with intuition: imagine the cross-section of each photon  as a disc of radius $\sqrt{2}/\sigma_n$. Starting from an initial condition of perfect superposition between the two discs, suppose to shift one disc with respect to the other by the amount  ${\delta r_{0n}}$.
Now, if ${\delta r_{0n}} \ll \sqrt{2}/\sigma_n$ the two discs have still a large superposition and the two photons remain largely indistinguishable. Vice versa, if ${\delta r_{0n}} \sim \sqrt{2}/\sigma_n$ the two discs separate completely and the superposition drops to zero. In this case the photons become ``quickly'' distinguishable. Analogous reasonings may be reproduced for the other DOFs.

Next, we consider the case of a non-factorable spectral amplitude, which couples wave vector coordinates $1$ and $2$.
Equation \eref{eq:Gaussian_f} still holds, but now $V$ has diagonal and non diagonal elements $ V_{nm} = \delta_{nm}/\sigma^2_n -(\delta_{n1} \delta_{m2}+\delta_{n2} \delta_{m1})/\sigma_{nm}$, where the real parameter $\sigma_{12}$ establishes the coupling, with  $\sigma_{12}^2 > \sigma_1^2 \sigma_2^2$ as required by positive definiteness of $V$. A straightforward calculation furnishes
\BEA\label{Corre}
R_{\sigma_n} = \frac{1}{2\sigma_n^2} \frac{1}{(1-\rho^2)^2}, \qquad (n=1,2),
\EEA
where $\rho \equiv \sigma_1 \sigma_2/\sigma_{12}$, with $\abs{\rho}<1$. If $\rho =0$ (uncoupled DOFs) we recover the results of  table  1. Vice versa, for increasing coupling one has $\rho \; \rightarrow \; 1$ and $R_{\sigma_n}$ grows unboundedly.
This result is of particular relevance to experimentalists: it tells us that wave packets whose cross section has the shape of an ellipse whose either major or minor axis does not lay on the plane of incidence (see figure 1), are much more sensitive to mode-mismatch than cylindrically symmetric wave packets. This occurrence strongly degrades photon indistinguishability and should be avoided, for example, in coalescence experiments \cite{Kuhn04,Specht09}.

Finally, we  examine the polarization DOFs of the two photons. Let us parameterize $\bi{p}^A$ and $\bi{p}^B$ as $\bi{p}^\lambda  = \left\{ \cos \vartheta^{\lambda} \exp \left( i \varphi^{\lambda}_{1} \right) ,  \, \sin \vartheta^{\lambda} \exp \left( i \varphi^{\lambda}_{2} \right),0 \right\}$,
with $\lambda = A,B$ and $|\bi{p}^A|=|\bi{p}^B|=1$. The results,  as  expansions in powers of $\sigma_1,\sigma_2$, are
\numparts
\BEA
R_{\vartheta} &\simeq 1+ \frac{\sigma_1^2-\sigma_2^2}{2k_0^2} \cos(2\vartheta) + \dots , \label{eqPoltheta} \\
%
%\EEA
%\BEA
R_{\varphi_n} &\simeq \frac{\sin^2(2\vartheta)}{4} \left[1+  \frac{\sigma_1^2-\sigma_2^2}{2k_0^2} \cos(2\vartheta) +   \dots \right], \label{eqPolphi}
\EEA
\endnumparts
with $n=1,2$. Unlike in the spectral and the scalar cases,  $R_{\vartheta}$ and $R_{\varphi_n}$ are dimensionless quantities.
From a physical point of view, this means that there is not a natural scale for the variation of the polarization DOFs. From equations (13) one sees that for astigmatic wave packets, namely for $\sigma_1 \neq \sigma_2$, there is a coupling between spectral and polarization DOFs that affects in equal manner both $R_{\vartheta}$ and $R_{\varphi_n}$. 
%
%\vspace{7.5truecm}
%
%%%%%%%%%%%%%%%%%%%%%%%%%%%%%%%%%%%%%%%%%%%%%%%%%%%%%%%%%%%%%%%%%%%%%%%%%%%%%%%%%%
%
%%%%%%%%%%%%%%%%%%%%%%%%%%%%%%%%%%%%%%%%%%%%%%%%%%%%%%%%%%%%%%%%%%%%%%%%%%%%%%%%%%%
%
 The term $(\sigma_1^2-\sigma_2^2)/{(2 k_0^2)}$ may be interpreted as a manifestation of the \emph{unavoidable} spin-orbit coupling occurring in transverse electromagnetic fields \cite{Bliokh11}. In addition, its absolute value furnishes the visibility of the coincidence fringes \cite{Padua10}.
Equation \eref{eqPolphi} shows that $R_{\varphi_n} \propto \sin^2(2\vartheta)$. As a consequence, for linearly polarized states with $2 \vartheta = 0, \pm \pi, \pm 2 \pi, \dots$, the phase is not a relevant DOF and  $R_{\varphi_n}=0$.

\begin{figure}[h!]
\begin{center}
\includegraphics[angle=0,width=12truecm]{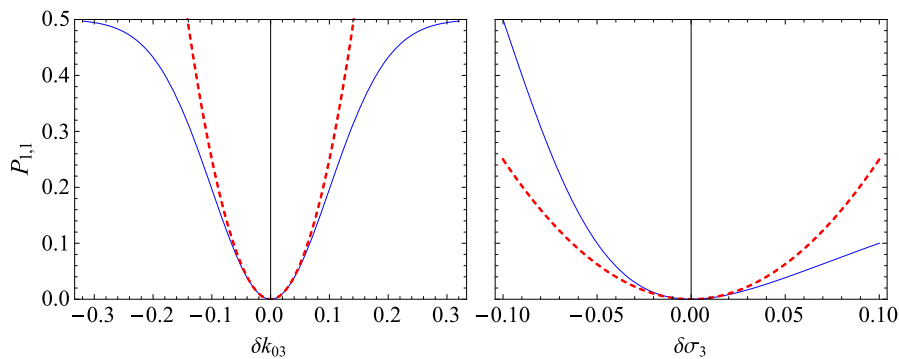}
\caption{\label{fig:2} (color online) Coincidence probability $P_{1,1}$ for $f = k_{03}$ (left) and $f = \sigma_{3}$ (right).
Blue lines: from equations \eref{eq:Gaussian_f} and \eref{eq100}; dashed red lines: $R_f \delta\!f^2/2$. In both plots we fixed $\sigma_3/k_{03} = 1/10$.}
\end{center}
\end{figure}

By definition, the rate of distinguishability $R_f[\psi]$ can be measured by interfering two independently created photons, each prepared in a state tunable in one specific DOF $f$. Such experiments have been realized for longitudinal spatiotemporal displacement $f=r_{03}$ \cite{Grangier06} and central frequency  $f=k_{03}$ \cite{Kuhn06}. Plotting the coincidence probability $P_{1,1}$ against the variation $\delta\!f$, yields a curve with a dip centered around $\delta\!f =0 $. By fitting this dip with a parabolic curve, as depicted in figure (\ref{fig:2}), one can straightforwardly extract $R_f[\psi]$ from the experimental data.
%
%%%%%%%%%%%%%%%%%%%%%%%%%%%%%%%%%%%%%%%%%%%%%%%%%%%%%%%%%%%%%%%%%%%%%%%%%%%%%%%%%%
%
%
%%%%%%%%%%%%%%%%%%%%%%%%%%%%%%%%%%%%%%%%%%%%%%%%%%%%%%%%%%%%%%%%%%%%%%%%%%%%%%%%%%%
%
%
\section{Mixed States}
In this section we consider the case occurring when both photons at the input ports of the BS are prepared in a statistical mixture.\footnote{We thank an anonymous Referee for suggesting us to consider the case of mixed states.} This is of great practical relevance especially for quantum information processing applications. We shall see that in certain circumstances there are profound differences for the rate of distinguishability of photons, as compared to the pure states case.  

For statistical mixtures equation \eref{eq10}, describing the probability of detecting the two photons at the two output ports $C$ and $D$, generalizes to:
\BEA\label{eqV10}
P_{1,1}\left[ \hat{\rho}^A, \hat{\rho}^B \right] =    \frac{1}{2} \left( 1 - \mathrm{Tr} \big[ \hat{\rho}^A \hat{\rho}^B\big] \right),
\EEA
where $\mathrm{Tr}[\dots] $ denotes the trace operation and $\hat{\rho}^A, \hat{\rho}^B$ are the normalized density operators describing the quantum state of photon $A$ and $B$ respectively. Differently from the pure state case where one has $P_{1,1}\left[ \psi, \psi \right]=0$, now 
\BEA\label{eqV20}
P_{1,1}\left[ \hat{\rho}, \hat{\rho} \right] =    \frac{1}{2} \left( 1 - \mathrm{Tr} \big[ \hat{\rho}^2 \big] \right),
\EEA
which is nonzero except for a pure state where $\mathrm{Tr} \big[ \hat{\rho}^2 \big]=\mathrm{Tr} \big[ \hat{\rho} \big]=1$. Now we proceed by analogy with the pure state case by choosing $\hat{\rho}^A = \hat{\rho}$ and 
\BEA\label{eqV30}
\hat{\rho}^B = \frac{\hat{\rho} + \delta     \hat{\rho}}{1 + \mathrm{Tr} \left[ \delta     \hat{\rho} \right]},
\EEA
where with $\delta     \hat{\rho}$ we denoted the (supposedly small) variation of $ \hat{\rho}$. By using \eref{eqV20} and \eref{eqV30} one can  calculate the difference $\Delta P_{1,1}\left[ \hat{\rho}\right] \equiv P_{1,1}\left[ \hat{\rho}, \frac{\hat{\rho} + \delta  \hat{\rho}}{1 + \mathrm{Tr} \left[ \delta  \hat{\rho} \right]}\right] - P_{1,1}\left[ \hat{\rho}, \hat{\rho}\right] $ as a perturbation expansion with respect to $\mathrm{Tr} [\delta     \hat{\rho}]$ by noting that for $\mathrm{Tr} [\delta \hat{\rho}]<1$ equation \eref{eqV30} may be written as a geometrical series: 
\BEA\label{eqV40}
\Delta P_{1,1}\left[ \hat{\rho}\right] = \frac{1}{2} \left( 
\mathrm{Tr} [ \hat{\rho}^2] \, \mathrm{Tr} [ \delta     \hat{\rho}] - \mathrm{Tr} [ \hat{\rho} \, \delta     \hat{\rho} ]
\right) \left(1 - \mathrm{Tr} [ \delta     \hat{\rho}]  + \mathrm{Tr} [ \delta     \hat{\rho}]^2  - \dots  \right) .
\EEA
Here an apparently striking difference with respect to the pure states case occurs: the first variation of $P_{1,1}$ is  linear in $\delta     \hat{\rho}$.
Moreover, if one chooses  $\delta     \hat{\rho}$ such that  $\mathrm{Tr} [\delta   \hat{\rho}]=0$ (we shall see later when this naturally occurs), then \eref{eqV40} reduces exactly to:
\BEA\label{eqV45}
\Delta P_{1,1}\left[ \hat{\rho}\right] = -\frac{1}{2}  \mathrm{Tr} [ \hat{\rho} \, \delta     \hat{\rho} ]
=  -\frac{1}{2} \left\langle \delta     \hat{\rho} \right\rangle.
\EEA
Moreover, it is not difficult to see that when the photons are prepared in pure states, so that one chooses $\hat{\rho}^A = |\Psi \rangle \langle \Psi |$, with $ \langle   \Psi |  \Psi \rangle = 1$, and 
\BEA\label{eqV50}
\hat{\rho}^B = \frac{\left( |\Psi \rangle + | \delta \Psi \rangle \right)\left( \langle \Psi | + \langle \delta \Psi| \right)}{ 1 + \langle \Psi | \delta \Psi \rangle + \langle \delta \Psi |\Psi \rangle  +  \langle  \delta  \Psi | \delta \Psi \rangle},
\EEA
then the first ``linear'' term on the right side of equation \eref{eqV40} becomes
\BEA\label{eqV60}
\frac{1}{2} \left( 
\mathrm{Tr} [ \hat{\rho}^2] \, \mathrm{Tr} [ \delta     \hat{\rho}] - \mathrm{Tr} [ \hat{\rho} \, \delta     \hat{\rho} ]
\right) &=  \frac{1}{2} \Big[ \langle \delta\Psi |\delta \Psi \rangle -  \left| \langle \Psi | \delta \Psi\rangle \right|^2 \Big]\nonumber\\
&=\frac{1}{2} \Big[ (\delta\psi ,\delta \psi ) -  \left| ( \psi , \delta \psi) \right|^2 \Big],
\EEA
which is clearly quadratic in $\delta \psi$ and we recover the results of section 2.

The second relevant difference between statistical mixtures and pure states is that in the first case we have at our disposal  also the parameters of the statistical distribution of the pure states (which constitute the  ensemble characterizing the photons) to yield the variation $\delta     \hat{\rho}$, in addition to the ``deterministic'' DOFs $f$ used previously. Specifically, we can distinguish amongst two different cases: given 
\BEA\label{eqV70}
\hat{\rho}^A = \sum_n w_n | \Phi_n(f) \rangle \langle \Phi_n(f) |, 
\EEA
with $w_n \geq 0$, $\sum_n w_n =1$ and $ \langle  \Phi_n(f) |  \Phi_m(f) \rangle = \delta_{nm}$, we can choose $\hat{\rho}^B$ either {\it a}) by varying the statistical distribution $w_n \rightarrow w_n + \delta w_n$, or {\it b}) by varying the DOFs $f$ of the wave packet, namely $\ket{\Phi_n(f)} \rightarrow \ket{\Phi_n(f + \delta\!f)}$.

{\it Case a}): Let
\BEA\label{eqV80}
\hat{\rho}^A = \sum_n w_n | \Phi_n \rangle \langle \Phi_n | \equiv \hat{\rho} \mathrm{~~and~~} \hat{\rho}^B = \sum_n \frac{w_n + \delta w_n}{1 + \sum_m \delta w_m} | \Phi_n \rangle \langle \Phi_n |, 
\EEA
represent the quantum states of photons $A$ and $B$, respectively. Then, a straightforward calculation yields, up to and including second order terms, 
\BEA\label{eqV90}
\Delta P_{1,1}\left[ \hat{\rho}\right] = \frac{1}{2} \sum_n \delta w_n \left( \mathrm{Tr} [\hat{\rho}^2] - w_n\right) \Bigl[ 1 -  
\sum_{m} \delta w_m   + \dots\Bigr] .
\EEA
Here, the first (linear) term is, in general, non zero.  A notable case occurs for $N$-dimensional maximally mixed states where $w_n = 1/N$ for all $n$ and $\mathrm{Tr} [\hat{\rho}^2] = 1/N \Rightarrow \Delta P_{1,1}\left[ \hat{\rho}\right] = 0 $. Physically, this means that photons prepared in maximally mixed states are intrinsically more robust against ``distinguishability'' than photons in pure states. However, the indistinguishability of maximally mixed states is per se very poor since for them $P_{1,1}[\hat{\rho},\hat{\rho}]=(1-1/N)/2$.

{\it Case b}): In this case we have
\numparts\label{eqV100}
\BEA
\hat{\rho}^A = & \; \sum_n w_n | \Phi_n(f) \rangle \langle \Phi_n(f) | \equiv \hat{\rho}(f), \\
\hat{\rho}^B = & \; \sum_n w_n | \Phi_n(f + \delta\!f) \rangle \langle \Phi_n (f + \delta\!f )| \equiv \hat{\rho}(f + \delta\!f), 
\EEA
\endnumparts
with $\delta     \hat{\rho} = \hat{\rho}^B - \hat{\rho}^A $ such that $\mathrm{Tr}[\delta     \hat{\rho}] =0$ as follows from normalization condition: $\langle\Phi_n(f) |\Phi_n(f)\rangle =1$ for all $f$ or, equivalently, $\mathrm{Tr}[\hat{\rho}(f )]=1=\mathrm{Tr}[\hat{\rho}(f + \delta\!f)]$.
The variation $\delta     \hat{\rho} =\hat{\rho}(f + \delta\!f)-\hat{\rho}(f ) $ can be written as a Taylor expansion
\BEA\label{eqV110}
\delta     \hat{\rho} = \delta\!f \frac{\partial \hat{\rho}}{\partial f} + \frac{\delta\!f^2}{2} \frac{\partial^2 \hat{\rho}}{\partial f^2} + \dots,
\EEA
and substituted into \eref{eqV45} to calculate
\BEA\label{eqV120}
\Delta P_{1,1}\left[ \hat{\rho}\right] =  -\frac{1}{2} \left\langle \delta     \hat{\rho} \right\rangle = -\frac{1}{2} \left[\delta\!f \left\langle \frac{\partial \hat{\rho}}{\partial f}\right\rangle + \frac{\delta\!f^2}{2} \left\langle \frac{\partial^2 \hat{\rho}}{\partial f^2}\right\rangle + \dots \right],
\EEA
where $\langle  \hat{O} \rangle$ denotes $\mathrm{Tr} [  \hat{\rho} \, \hat{O} ] $.
By using the evident relation  
\BEA\label{eqV130}
\frac{\partial }{\partial f} \langle \Phi_n(f) | \Phi_n(f) \rangle=0,
\EEA
 it is not difficult to prove that 
\BEA\label{eqV140}
\left\langle \frac{\partial \hat{\rho}}{\partial f}\right\rangle =0.
\EEA
Thus, \eref{eqV120} can be rewritten as
\BEA\label{eqV150}
\Delta P_{1,1}\left[ \hat{\rho}\right] =  -\frac{\delta\!f^2}{4} \left\langle \frac{\partial^2 \hat{\rho}}{\partial f^2}\right\rangle + \dots.
\EEA
Equation \eref{eqV150} shows that for case {\it b}), namely when we vary one DOF of the photon state, say $f$, the variation $\Delta P_{1,1}$ of the probability coincidence is at least of the second order with respect to $\delta\!f$. This result not only reproduces our findings in section 2, but extend their validity to the case of mixed states. Of course, \eref{eqV150} is also valid for pure states and, therefore, we can rewrite the rate of distinguishability as proportional to the expectation value of the operator $\partial^2 \hat{\rho}/\partial f^2$:
\BEA\label{eqV155}
R_f\left[ \hat{\rho} \right] =  -\frac{1}{2} \left\langle \frac{\partial^2 \hat{\rho}}{\partial f^2}\right\rangle.
\EEA

Let us apply the results obtained above to the exactly tractable case of a \emph{partially polarized} paraxial beam of light prepared in a well defined spatial mode decoupled from polarization DOFs. The relevant single-photon density operator is given by
\BEA\label{eqV160}
\hat{\rho}=\hat{\rho}(\alpha,\vartheta, \varphi) = \cos^2 \alpha |\Psi \rangle \langle \Psi | + \sin^2 \alpha |\Psi_\perp \rangle \langle \Psi_\perp |,
\EEA
where $\alpha\in[0,2\pi[$ and
\BEA\label{eqV170}
|\Psi \rangle = \cos \vartheta |x \rangle + \sin \vartheta \, e^{i \varphi}|y \rangle \mathrm{~~and~~}  |\Psi_\perp \rangle = - \sin \vartheta \, e^{-i \varphi} |x \rangle + \cos \vartheta |y \rangle,
\EEA
with $| x \rangle$ and $| y \rangle$ representing two normalized orthogonal polarization states: $\langle x | y \rangle = 0$. According to the preceding analysis, we can study two different cases: 
\BEA\label{eqV180}
\mathrm{{\it a})} \quad \left\{ 
\begin{array}{ll}
	\hat{\rho}^A &= \hat{\rho}(\alpha,\vartheta, \varphi), \\
	\hat{\rho}^B &= \hat{\rho}(\alpha + \delta \alpha,\vartheta, \varphi), \\
\end{array}
 \right.
 \qquad
 \mathrm{ {\it b})} \quad \left\{ 
\begin{array}{ll}
	\hat{\rho}^A &= \hat{\rho}(\alpha,\vartheta, \varphi), \\
	\hat{\rho}^B &= \hat{\rho}(\alpha,\vartheta + \delta \vartheta, \varphi). \\
\end{array}
 \right. 
\EEA

For case {\it a}) a straightforward calculation furnishes
\BEA\label{eqV190}
\Delta P_{1,1}\left[ \hat{\rho}\right] &= \frac{1}{4} \sin( \delta \alpha) \left[ \sin (\delta \alpha ) + \sin(4 \alpha + \delta \alpha) \right] \nonumber \\
&=\frac{\delta \alpha}{4}   \sin( 4 \alpha)  + \frac{\delta \alpha^2}{2}  \cos^2 (2 \alpha) + \dots.
\EEA
Equation \eref{eqV190} shows that as a consequence of the variation of the parameter $\alpha$ defining the statistical distribution of the mixed state, the corresponding variation of $\Delta P_{1,1}$ is linear in $\delta \alpha$ and the ``quadratic'' rate of distinguishability cannot be defined here. However, it should be noticed that our rate of distinguishability coincides with  the second order coefficient of the Taylor expansion of $\Delta P_{1,1}$. Therefore, in principle, if one identifies the $n$th order coefficient of such expansion with the $n$th rate of distinguishability $R_f^{(n)}$ (with $R_f=R_f^{(2)}$) a hierarchy between $n$th and $(n+1)$th rate of distinguishability is unambiguously established. Thus, for example, in the case above the existence of the first order rate of distinguishability $R_f^{(1)}=\sin(4\alpha)/4$ indicates a greater attitude of photons to become distinguishable when there statistical distribution is varied.
Note that, as expected, for the maximally mixed state attained at $\alpha = \pi/4$ one has \emph{exactly} $\Delta P_{1,1}\left[ \hat{\rho}\right] =0$, as previously found on the ground of general considerations.

For case {\it b}) we put $f = \vartheta$ and obtain
\BEA\label{eqV200}
\Delta P_{1,1}\left[ \hat{\rho}\right] &= \frac{1}{2} \cos^2(2 \alpha) \sin^2(\delta \vartheta)\nonumber \\
& = \frac{\delta \vartheta^2}{2}  \cos^2 (2 \alpha) + \dots.
\EEA
By an explicit calculation one can see that \eref{eqV200} is in perfect agreement with  \eref{eqV155}. Moreover, once again, for $\alpha = \pi/4$ one retrieve the expected result $\Delta P_{1,1}\left[ \hat{\rho}\right] =0$. For pure states occurring at $\alpha\in\{0,\pi/2,\pi,3\pi/2\}$ this result does also coincide with \eref{eqPoltheta} which furnishes $R_\vartheta=1$ in absence of spin-orbit coupling.

\section{Conclusions}
In this work we have investigated photon distinguishability from an operational point of view. 
We introduced a new parameter, the rate of distinguishability $R_f[\psi]$, which furnishes a quantitative measure of the distinguishability of photons (prepared in the state $|\Psi \rangle$) with respect to the DOF $f$. 
Our main results are summarized by Equations (\ref{eq20},\ref{eq60},\ref{Corre},\ref{eqV155}), and  table  1. In particular,  \eref{Corre} quantifies the degradation of photon distinguishability 
due to coupling between different DOFs. Moreover, we extended the definition of $R_f[\psi]$ form the pure state $\ket{\Psi}$ to the density operator $\hat{\rho}$. For this case we found that the variation of the statistical distribution of the incoming photons affects their degree of distinguishability which is, in practice, increased. As a final remark, we stress that  $R_f[\psi]$ is  experimentally accessible via the measurement of the two-photon coincidence probability. 
\end{document}